# Advection-Dominated Accretion Model for Sagittarius A*: Evidence for a $10^6 M_\odot$ Black Hole at the Galactic Center

Ramesh Narayan, Insu Yi & Rohan Mahadevan
Harvard-Smithsonian Center for Astrophysics
60 Garden St, Cambridge, MA 02138, USA

The enigmatic radio source Sagittarius A* (Sgr A* ) at the centre of our Galaxy appears to be a low-luminosity version of active galactic nuclei (AGN) in other galaxies[1]. By analogy with AGN models[2], it has been proposed that Sgr A* may be a massive accreting black hole[1,3]. This is apparently confirmed by dynamical evidence that the center of the Galaxy has a dark object with a mass $\lesssim 10^{6.5} M_\odot$ which seems to have a gravitational influence on gas and stellar motions out to $\sim 1$ pc[1,4,5]. The black hole hypothesis is, however, problematical because no model of Sgr A* has been able to explain the observed spectrum in any self-consistent way, and there is no consensus on either the mass of the black hole or the mass accretion rate[6-8]. Sgr A* has been observed in the radio[9-11], sub-mm[12,13], infrared[12-15], and X-ray[16-18] bands, and the various detections and flux upper limits cover more than ten decades of frequency, from $\nu \lesssim 10^9$ Hz (meter wavelength radio band) up to $\nu \gtrsim 10^{19}$ Hz (150 keV X-rays). We present a robust model of Sgr A* in which a $10^6 M_\odot$ black hole accretes at the rate of a few $\times 10^{-6} M_\odot \mathrm{yr}^{-1}$. This model fits the entire spectrum self-consistently. The unique feature of the model is that the flow is advection-dominated, i.e. most of the energy which is viscously dissipated in the differentially rotating flow is carried along with the gas and lost through the horizon. The apparent success of this model in explaining the data may be considered "proof" that horizons are real and that a massive black hole does exist at the Galactic center.

Figure 1 shows the spectral data on Sgr A*[9-18]. The bulk of the emission occurs in sub-mm at $\nu \sim 10^{12}$ Hz and in near infrared at $\nu \sim 10^{14} - 10^{14.5}$ Hz, but between these two bands there is a pronounced and very significant dip in the luminosity in the far IR band[12,13]. (Note that the $10^{14.5}$ Hz flux determination[15] is from a crowded field and is better interpreted as an upper limit.) Sgr A* has also been detected as a variable soft X-ray source (3–30 keV, variable by a factor of 6 [16,17]) and there are recent upper limits[18] on its luminosity in two hard X-ray bands (35–75 and 75–150 keV). There is no direct information in optical and UV because of obscuration[1]; however, the maximum luminosity in these bands is limited to $< 3 \times 10^{39} \mathrm{ergs}^{-1}$.[1,8,11]

If Sgr A* is to be understood as an accretion powered system, it is necessary to model the viscous flow of the accreting gas as well as the radiation mechanisms whereby the dissipated energy is converted into the radiation we observe. The most commonly adopted paradigm in this field is the so-called thin accretion disk model[2,8] which assumes that the cooling is efficient and that the energy released through viscosity is radiated immediately. This leads to a number of simplifications in the dynamics of the flow and in the calculation of the emission. However, the model does not apply to Sgr A* since it predicts a spectrum[2] which is very different from the data shown in Fig. 1.

Since the thin disk model makes approximations which may not always be valid, some authors have expanded the model to include the effects of radial pressure gradients and radial energy transport. The extended theory is referred to as the slim disk model[19] and one of its features is that it allows energy advection, i.e. the gas is free to transport some of the viscously dissipated energy as stored entropy. The fraction $f$ of the energy which is advected is solved for self-consistently as a function of radius $R$. The slim disk equations have been successful in the study of disk instabilities[19,20] and in modeling boundary layers of accretion disks around white dwarfs and pre-main sequence stars[21,22].

The present work grew out of investigations of *advection-dominated* flows[23-26], where energy advection is not a perturbation, or included merely for self-consistency, but actually dominates the physics of the flow. In terms of the parameter $f$, we consider accretion flows where $f \to 1$, and the local radiative efficiency $1 - f \ll 1$. In the case of accretion onto a black hole, advection-dominated flows are possible[25] whenever the mass accretion rate satisfies $\dot{M} \lesssim 10^{-2} - 10^{-1} \dot{M}_{Edd}$, where $\dot{M}_{Edd} = 2.2 \times 10^{-8} M M_\odot \mathrm{yr}^{-1}$ is the Eddington accretion rate and $M$ is the mass of the accreting object in solar mass units. The model of Sgr A* presented here corresponds to $\dot{M} \sim 10^{-4} \dot{M}_{Edd}$.

The basic framework of our low-$\dot{M}$ advection-dominated models is described in ref. 25. Because the cooling is inefficient ($1-f \ll 1$), these models have a very high ion temperature. The ion pressure therefore becomes dynamically important, causing the gas to be almost spherical in morphology and to rotate slowly. The model thus represents a hybrid between a so-called "ion torus"[27] and a non-rotating spherical accretion flow[28,6,7], except that we solve self-consistently for the advection of energy and the transport of angular momentum. We also compute the thermal structure of the accreting gas more completely and obtain realistic spectra.

We model viscosity by the usual $\alpha$ prescription[29], where the kinematic viscosity coefficient is taken to be $\alpha c_s^2/\Omega_K$, $c_s$ being the local sound speed and $\Omega_K$ the Keplerian angular velocity. This simple prescription can plausibly account for the effects of hydrodynamic turbulence, convection[23,24], and magnetic stresses. We assume that the accreting plasma is composed of gas and magnetic fields, such that a fraction $\beta$ of the pressure is supplied by the gas and a fraction $1-\beta$ by the fields. We take $\alpha$, $\beta$ to be independent of $R$. We set $\alpha = 0.3$ (as in our previous work[25]) and discuss two values of $\beta$, viz. 0.5 (equipartition between gas and magnetic pressure) and 0.95 (gas pressure dominant). The parameters $\alpha$ and $\beta$, along with the mass $M$ and accretion rate $\dot{M}$, complete the specification of a model.

Given these parameters, and using a self-similar solution[23] as the local representation of the flow, we solve[25] for all the properties of the accreting gas, such as the density $\rho$, angular velocity $\Omega$, radial velocity $v_R$, sound speed $c_s$, etc., at each radius $R$. We determine the ion and electron temperatures, $T_i$ and $T_e$, self-consistently by solving for energy balance among various processes like viscous heating, advection, energy transfer from ions to electrons (via Coulomb collisions), and radiative cooling of the electrons. For the cooling we include synchrotron emission, bremsstrahlung, and Comptonization. One simplification is that we use a Newtonian model of gravity down to the Schwarzschild radius, $R_S = 2GM/c^2$, and assume that the gas disappears through the black hole horizon at $R = R_S$. Also, we do not model the details of the sonic transition[28] near the horizon or the gravitational redshift of the escaping radiation. These approximations will doubtlessly lead to quantitative errors in the results but we believe that the model captures most of the esssential physics.

The solid lines in Fig. 2 show the variations of $T_i$, $T_e$ and $1-f$ as functions of $R$ for one of our models, with $\alpha = 0.3$, $\beta = 0.5$, $M = 10^6 M_\odot$ and $\dot{M} = 1.3 \times 10^{-4} \dot{M}_{Edd} = 2.9 \times 10^{-6} M_\odot \text{yr}^{-1}$. At large radii, $T_i$ and $T_e$ are both very nearly equal to the local virial temperature, $T_{vir} \sim 2 \times 10^{12} \beta (R_S/R)$ K. Energy transfer from ions to electrons is very efficient here and so the two species come into thermal equilibrium. However, the cooling of the electrons is inefficient, and the flow is therefore advection-dominated. For $R \lesssim 10^2 R_S$, the electron temperature saturates at $T_e \sim 10^{10}$ K but the ion temperature continues to track $T_{vir}$. Although electron cooling is efficient at these temperatures, the ion-electron coupling becomes weak, and once again the flow is advection-dominated. Thus, at all radii, the radiative efficiency $1-f \ll 1$.

The solid line in Fig. 1 shows the spectrum predicted by this model. Basically, there are four peaks in $\nu L(\nu)$, where $L(\nu)d\nu$ is the power radiated by the source between frequencies $\nu$ and $\nu + d\nu$. The peak labeled S is due to synchrotron emission by the thermal electrons in the magnetic field of the plasma, the peaks C1 and C2 are due to Comptonization[30] (single scattering and double scattering) of the synchrotron radiation by the hot electrons, and the peak B is due to bremsstrahlung radiation. The presence of these four peaks is a robust feature of all low-$\dot{M}$ advection-dominated models, but the frequencies at which the peaks appear and their relative heights depend on the parameters.

The position of the synchrotron peak S depends primarily on $M$, while its height varies roughly as $\dot{M}$. Thus, merely by fitting the sub-mm data and requiring that the model satisfy the far-IR limits at $\nu \sim 10^{13} - 10^{14}$ Hz, it is possible to estimate $M$ and $\dot{M}$ fairly well. The sharp drop at the high-$\nu$ end of the S peak arises because most of the luminosity is emitted at the innermost radius where the synchrotron emission is highly self-absorbed. Allowing for fluctuations in the magnetic field strength and gravitational redshift effects, and modeling the spectrum more accurately above the self-absorption limit, will round the peak and produce a more gradual fall-off, but a significant dip between S and C1 is always expected.

At radio frequencies, $\nu < 10^{10}$ Hz, the model spectrum lies below the observations. This emission comes from large radii where the gas temperature may conceivably be modified by radiative transfer or other transport effects which are not included adequately in the model. Another possibility is that some of the accreting gas may undergo shocks and participate in an outflow, as advection-dominated flows seem prone to do[23,24], and this gas may radiate in radio waves through shock-accelerated non-thermal electrons. In any case, the luminosity at these frequencies is very low and the discrepancy is probably not serious for the overall viability of the model.

Because the synchrotron emission is highly self-absorbed, the emission at each $\nu$ in the S peak originates essentially at a single radius. The radiation at 86 GHz, for instance, comes from $R = 57 R_S = 1.1$ AU. This is in excellent agreement with a recent measurement[31,32] which gives a size $\sim 1.1$ AU.

The C1 and C2 peaks are obtained by single and double Compton scattering of the synchrotron photons. The frequency shift between the S and C1 peaks and that between C1 and C2 are determined primarily by $T_e$ since the mean energy boost $\langle A \rangle$ of a photon in each Compton scattering depends on the distribution of Lorentz $\gamma$ of the scattering electrons. In the models shown here, the maximum electron temperature is $T_{e,max} = 8 - 14 \times 10^9$ K which gives $\langle A \rangle \sim 30 - 100$. Note that the electron scattering optical depth is very small, $\tau_{es} < 10^{-2}$. Therefore, most of the synchrotron photons escape and only a small fraction $\sim \tau_{es}$ is Compton-scattered; the fraction that is scattered twice is $\sim \tau_{es}^2$. The heights of the C1 and C2 peaks relative to S are therefore proportional to $\langle A \rangle \tau_{es}$ and $\langle A \rangle^2 \tau_{es}^2$ respectively. The flux in the S peak is $\propto \dot{M}$, as is the optical depth $\tau_{es}$. Therefore, the flux in the C1 peak varies as $\sim \dot{M}^2$ and that in C2 as $\sim \dot{M}^3$. If the source has a variable $\dot{M}$ for any reason, these peaks will vary in amplitude, with the largest fluctuations expected in C2. Indeed, the luminosity of Sgr A* in the soft X-ray band is known to be variable[16]. The observed fluctuations can be explained by changes in $\dot{M}$ by a few tens of percent over timescales of years. Variability may also arise from fluctuations in the parameter $\beta$, due to random fluctuations in the magnetic field strength. The near-IR signal in the C1 peak should show lower amplitude variations; we are not aware of any measurements. Since the S, C1 and C2 peaks are closely coupled in the model, variations in their heights should be correlated in time, a testable prediction.

The bremsstrahlung peak B occurs at the thermal frequency, $\nu = kT_e/h$, so its position is determined primarily by $T_e$. Figure 1 suggests that the model has this peak too close to the hard X-ray limits. With a higher $T_e$ the peak will move to a higher $\nu$ and the discrepancy can be removed. Gamma-ray observations of Sgr A* at energies $\sim 1$ MeV will be able to constrain the properties of the bremsstrahlung emission and through this $T_e$. The height of the B peak varies as $\dot{M}^2$ since bremsstrahlung, being a two-particle process, is proportional to $\rho^2$.

We briefly mention the effect of the parameters $\alpha$ and $\beta$. For $\alpha < 1$, we find that models with a given value of $\dot{M}/\alpha$ have virtually identical spectra. Therefore, in Fig. 1, any change in $\alpha$ merely requires rescaling $\dot{M}$ by the same factor. The effect of $\beta$ is more interesting. As we increase $\beta$, i.e. as we increase the importance of gas pressure relative to magnetic pressure, we find that $T_e$ goes up. The dashed lines in Figs. 1 and 2 show an example with $\beta = 0.95$. This model is perhaps a better fit to the data than the $\beta = 0.5$ model. In this case, $T_{e,max} = 1.4 \times 10^{10}$ K and the source size at 86 GHz is 0.9 AU.

We should emphasize that there is very little fine tuning in the models. Merely by fitting the S peak we are able to estimate $M$ and $\dot{M}$, and with these parameters we obtain a more or less satisfactory fit to the rest of the spectrum. The dip between S and C1 is a robust prediction of the model which nicely explains a feature in the data very difficult to fit with other models. Further, we must emphasize that the model does not invoke a number of different radiating regions. All the emission comes from the same hot electrons (over a range of $R$), and the different peaks in the spectrum are just signatures of different radiation mechanisms. Considering how simple the model is in its basic structure, the level of agreement with the data is rather good, indicating perhaps that the model is correct in its essentials.

The accreting material reaches relativistic temperatures in our model because it falls into a relativistic potential. The only objects known in astrophysics with such deep potentials are neutron stars and black holes, and of the two, neutron stars are believed to be limited to masses $\lesssim 3 M_\odot$. Since the accreting object in our model has a mass of $10^6 M_\odot$, this implies that Sgr A* must be a black hole.

While such an argument based on mass is quite plausible and is routinely employed in searches for black holes, we note that the crucial feature which distinguishes a black hole from other objects is not mass but rather the existence of a *horizon*, a surface which separates the interior of the black hole from the rest of the universe. The evidence for a black hole would be much more secure if one could prove that the object in question has a horizon into which matter can fall but out of which nothing, not even light, can escape. Remarkably, we believe that our model of Sgr A* comes close to supplying such a proof.

Recall that the most important aspect of our model is that it is advection-dominated. This is the reason that we are able to invoke a fairly large accretion rate, $\dot{M} \sim$ few $\times 10^{-6} M_\odot \, \text{yr}^{-1}$, and yet have a total luminosity as low as $\sim 2 \times 10^{37} \text{ergs}^{-1}$, much less than $\dot{M}c^2 \sim 10^{41} \text{ergs}^{-1}$. Where does the rest of the energy go? If Sgr A* were a normal object, then all the advected energy would finally be radiated from its surface. This energy will presumably come out somewhere in the electromagnetic band and will have a net luminosity $\sim$ few $\times 10^{40} \text{ergs}^{-1}$. There is no evidence at all that Sgr A* is emitting this much radiation. In contrast, in our model of Sgr A*, something like 99.9% of the energy is advected with the gas and disappears through the horizon and only 0.1% is radiated (see $1-f$ in Fig. 2). It is this feature above all else that allows the model to work. Therefore, if one accepts the model we have proposed, then one automatically has to accept the fact that the central object in Sgr A* actually does have a horizon. To our knowledge this is the first "proof" of the reality of black hole horizons. Unfortunately, the proof is not quite complete because, in principle, the missing energy could be emitted in the form of kinetic energy in an outflow. Although there is no evidence, as far as we know, for such an outflow, it is probably hard to rule one out.

The model we have presented for Sgr A* can be applied to other black hole candidates accreting at low rates. The majority of ultra-soft X-ray transients (also known as X-ray novae) in the Galaxy are believed to be stellar-mass black holes ($M \sim 10 M_\odot$)[33]. In outburst, these sources accrete at close to $\dot{M}_{Edd}$. Under these conditions, cooling is efficient, the accreting gas probably forms a thin accretion disk, and the total luminosity is expected to be $\sim 0.1 \dot{M}c^2$. However, in quiescence, the X-ray transients have $\dot{M} \ll \dot{M}_{Edd}$, and in this phase the sources are possibly in an advection-dominated state. We ought to be able to fit the spectra of these systems with the same model we have proposed for Sgr A* but with different choices of $M$ and $\dot{M}$. AGN in their less active mode should again be similar to Sgr A*, though with larger masses, $M \sim 10^6 - 10^9 M_\odot$. It has been suggested that neighbouring galaxies like M31 and M32 harbour supermassive black holes[34,35]. If these black holes are accreting at all, they are probably doing so in an advection-domination state and they should have weak radio and sub-mm emission corresponding to the S peak. The C1, C2 and B peaks in the spectrum will also be present, but since they depend more sensitively on $\dot{M}$ they will be undetectable unless $\dot{M}$ is quite high.

Among Galactic X-ray binaries, it has turned out to be difficult to distinguish between black hole and neutron star systems because at high $\dot{M}$ both have similar luminosities $\sim 0.1 \dot{M}c^2$. However, at low $\dot{M}$, if the accretion occurs in an advection-dominated mode, the difference between an object with a horizon and one with a surface is quite large. In the case of a low $\dot{M}$ accreting neutron star we expect that, even if the accretion flow is advection-dominated, all the gravitational energy released in the accretion will eventually be radiated from the stellar surface, and the total luminosity will continue to be $\sim 0.1 \dot{M}c^2$. In contrast, a black hole in a similar situation will release a far smaller fraction of the rest energy of the accreting material. This distinction between neutron star and black hole systems could perhaps be used to discover new black holes in the Galaxy.

We thank J. E. Grindlay for useful comments.

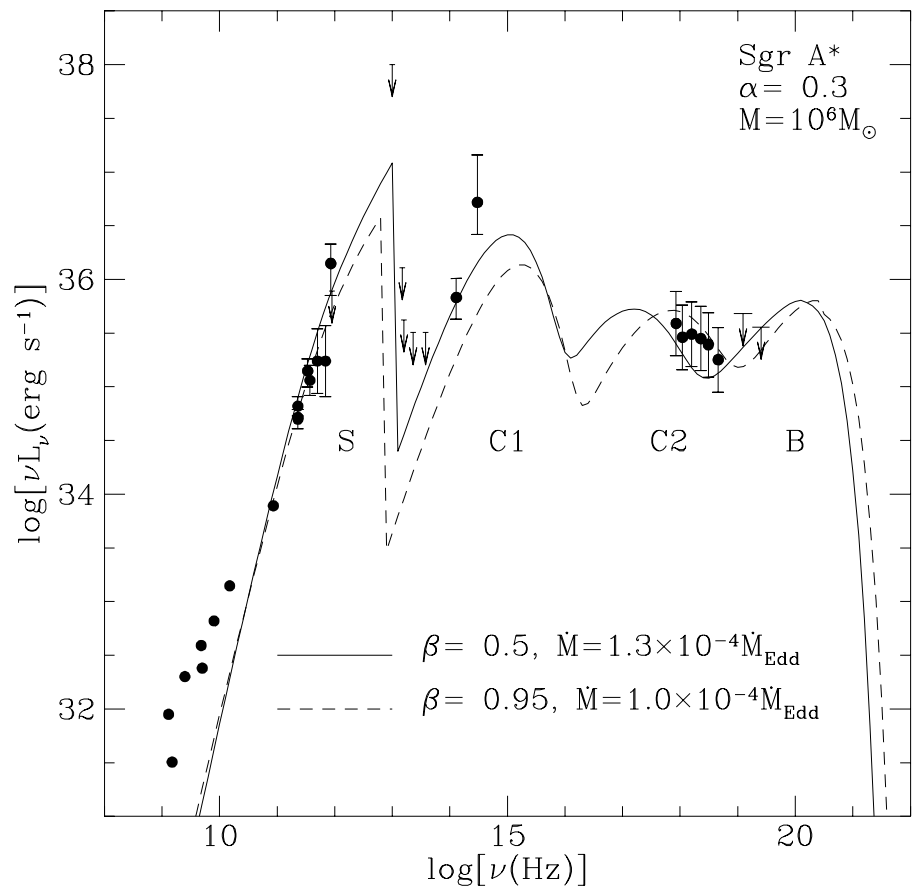

Figure 1. Filled circles indicate measurements of the spectrum of Sgr A* at various frequencies. The arrows represent upper limits. The solid and dashed lines show the spectra corresponding to two models described in the paper.

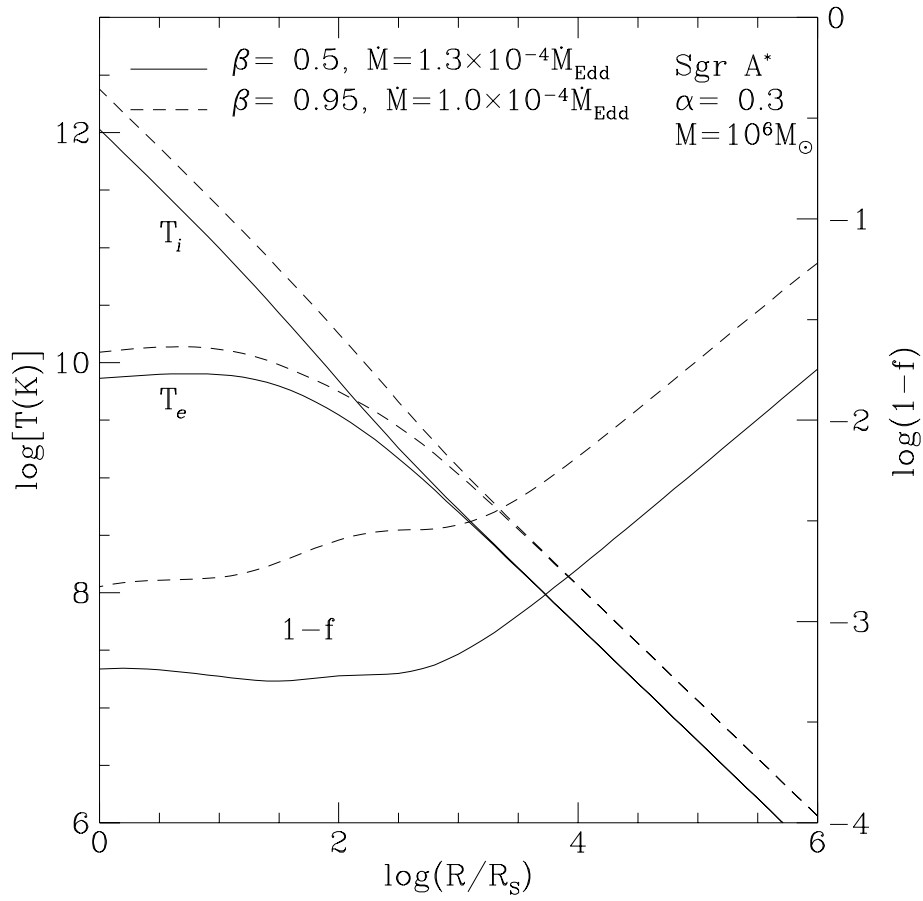

Figure 2. The variation with radius of the ion temperature, $T_i$, the electron temperature, $T_e$, and the radiative efficiency, $1 - f$, for the two models shown in Fig. 1.